\documentclass[a4paper, 12pt]{article}

\usepackage{amsmath}
\usepackage{graphicx}

\begin{document}
\renewcommand{\theequation}{\thesection .\arabic{equation}}

\newcommand{\sign}{\operatorname{sign}}
\newcommand{\Ci}{\operatorname{Ci}}
\newcommand{\tr}{\operatorname{tr}}

\newcommand{\beq}{\begin{equation}}
\newcommand{\eeq}{\end{equation}}
\newcommand{\beqn}{\begin{eqnarray}}
\newcommand{\eeqn}{\end{eqnarray}}

\newcommand{\slp}{\raise.15ex\hbox{$/$}\kern-.57em\hbox{$ \partial $}}
\newcommand{\lnA}{\raise.15ex\hbox{$/$}\kern-.57em\hbox{$A$}}
\newcommand{\unmedio}{{\scriptstyle\frac{1}{2}}}
\newcommand{\uncuarto}{{\scriptstyle\frac{1}{4}}}

\newcommand{\trial}{_{\text{trial}}}
\newcommand{\true}{_{\text{true}}}
\newcommand{\const}{\text{const}}

\newcommand{\intp}{\int\frac{d^2p}{(2\pi)^2}\,}
\newcommand{\intx}{\int_C d^2x\,}
\newcommand{\inty}{\int_C d^2y\,}
\newcommand{\intxy}{\int_C d^2x\,d^2y\,}

\newcommand{\bP}{\bar{\Psi}}
\newcommand{\bc}{\bar{\chi}}
\newcommand{\hs}{\hspace*{0.6cm}}

\newcommand{\bra}{\left\langle}
\newcommand{\ket}{\right\rangle}
\newcommand{\bracket}{\left\langle\,\right\rangle}

\newcommand{\D}{\mbox{$\mathcal{D}$}}
\newcommand{\N}{\mbox{$\mathcal{N}$}}
\newcommand{\Lag}{\mbox{$\mathcal{L}$}}
\newcommand{\V}{\mbox{$\mathcal{V}$}}
\newcommand{\Z}{\mbox{$\mathcal{Z}$}}
\newcommand{\A}{\mbox{$\mathcal{A}$}}
\newcommand{\B}{\mbox{$\mathcal{B}$}}
\newcommand{\C}{\mbox{$\mathcal{C}$}}
\newcommand{\E}{\mbox{$\mathcal{E}$}}


\vspace{2cm}

\begin{center}

{\Large {\bf Functional bosonization with time dependent
perturbations}}

\vspace{1.3cm}

Carlos M. Na\'on$^{a,b}$, Mariano J. Salvay$^{a,b}$, Marta L.
Trobo$^{a,b}$ \footnote{e-mail: salvay@fisica.unlp.edu.ar,
naon@fisica.unlp.edu.ar, trobo@fisica.unlp.edu.ar}

\vspace{.8cm}

$^a$ {\it Instituto de F\'{\i}sica La Plata, Departamento de
F\'{\i}sica, Facultad de Ciencias Exactas, Universidad Nacional de
La Plata.  CC 67, 1900 La Plata, Argentina.}

\smallskip

$^b$ {\it Consejo Nacional de Investigaciones Cient\'{\i}ficas y
T\'ecnicas, Argentina.}

\vspace{1cm}

\begin{abstract}
We extend a path-integral approach to bosonization previously
developed in the framework of equilibrium Quantum Field Theories,
to the case in which time-dependent interactions are taken into
account. In particular we consider a non covariant version of the
Thirring model in the presence of a dynamic barrier at zero
temperature. By using the Closed Time Path (Schwinger-Keldysh)
formalism, we compute the Green's function and the Total Energy
Density of the system. Since our model contains the Tomonaga
Luttinger model as a particular case, we make contact with recent
results on non-equilibrium electronic systems.
\end{abstract}
\end{center}

\vspace{1 cm}


\noindent{\it Pacs:} 11.10.Lm, 05.30.Fk


\newpage
\section{Introduction}

The study of a quantum field theory involving time-dependent
perturbations is a relevant issue for a large variety of phenomena
such as formation and growth of domains in high energy physics
\cite{Boyanovsky}, decoherence and photon pair creation in a
homogeneous electric field \cite{Cooper}, electron transport in
solids \cite{Mahan}, etc. In recent years there has been an
intense activity focused on the analysis of non equilibrium
situations in the context of low dimensional electronic systems
\cite{Chamon}. In particular, theoretical work on one dimensional
(1D) structures relies heavily on field theories obtained as
continuum limits of original lattices and on the systematic use of
bosonization and renormalization group methods \cite{Reviews}. In
this context, an alternative path-integral bosonization method,
valid for the description of systems in equilibrium, was first
established in \cite{Nosotros}. In view of the interesting open
problems in 1D non-equilibrium fermionic systems, in this work we
generalize the above mentioned functional bosonization method to
account for the time evolution of the system. To this end we
consider a generalized (non covariant) Thirring like model in the
presence of a dynamic barrier at zero temperature. By using the
Closed Time Path (Schwinger-Keldysh) formalism (CTP)\cite{CTP}, we
compute the Green's function and the Total Energy Density (TED) of
the system. Our model contains the Tomonaga-Luttinger model
(paradigm of Luttinger liquid behavior) as a particular case
\cite{g-ology}. This allows us to make contact with recent
interesting results on non-equilibrium many-fermion electronic
systems \cite{Komnik} which were obtained through operational
bosonization. These authors were able to obtained the TED for a
generic impurity (without specifying its spatial dependence) by
fixing from the beginning a spatial point of interest and
disregarding the adiabatic switching of the time-dependent
interaction. Although this procedure is fully justified on
physical grounds, taking into account the temporal step function
brings additional difficulties to the calculations and it is not
evident a priori that results should coincide, unless a large time
limit is considered. Then, as a secondary task, we have taken into
account the influence of the adiabatic switching, at least at the
level of the Green's function. This, in turn, prevented us from
obtaining a TED for an arbitrary impurity geometry. We were then
forced to specify a particular barrier, but, on the other hand, we
could get the spatial dependence of the energy distribution.

The paper is organized as follows. In Section 2 we show how to
conciliate the path-integral approach to bosonization based on
decoupling changes in the functional measure with the procedures
of the Closed Time Path formulation. In Section 3 we compute the
Green's functions and specialize our results to the case of an
adiabatically switched barrier of width $a$ and height $\lambda$
which oscillates with frequency $\Omega$. In Section 4 we compute
the Total Energy Distribution function and its temporal average,
in the large distance (time) approximation. This enables us to
analyze the behavior of this distribution as function of energy,
strength of the dynamical barrier and of the electron-electron
interactions. We also comment on the spatial distribution for
given energies. Our results are in qualitative agreement with the
findings of ref. \cite{Komnik}, which were obtained by using
conventional (operational) bosonization.  In Section 5 we
summarize our results and gather our conclusions.

\section{CTP Formalism and Functional Bosonization}
\setcounter{equation}{0}

 In the S matrix formalism the time is taken
over the interval $(-\infty, +\infty)$. The state at $t= -\infty$
is well defined as the ground state of the non-interacting system.
The interactions are turned on slowly. At $t \approx 0$ the fully
interacting ground state is $\psi(0) = S(0, -\infty)\,\Phi_0$. In
condensed matter physics the state at $t \rightarrow \infty$ must
be defined carefully. If the interactions remain on, then this
state is not well described by the non-interacting ground state
$\Phi_0$. Alternately, one could require that the interaction turn
off at large times, which returns the system to the ground state
$\Phi_0$. As it is well known, in 1961 Schwinger \cite{CTP}
suggested another method of handling the asymptotic limit $t
\rightarrow \infty$. He proposed that the time integral in the S
matrix could be separated in two pieces: one going from $-\infty$
to a certain time $\tau)$ and the second one going from $\tau$ to
$-\infty$ (eventually one could also let $\tau \rightarrow
\infty$). Thus the integration path is a time loop C, which starts
and ends at $t = -\infty$, and it is composed by two branches
$C_+$ ($-\infty,\tau$) and $C_-$ ($\tau,-\infty$). The advantage
of this method is that one starts and ends the S matrix expansion
with a known state $ \psi(-\infty) = \Phi_0$, which usually is the
only ground state one knows exactly.
 For equilibrium phenomena the time loop
method of evaluating the S matrix gives results that are identical
to the other methods. Perhaps its main practical advantage resides
in the possibility of describing non-equilibrium phenomena using
Green's functions. The equation of motion for the Green's function
can be cast into the form of a quantum Boltzmann equation for the
transport theory \cite{Mahan}. A disadvantage of the time-loop
method is that it employs four different Green's functions. They
are all correlation functions which relate the field operator of
the particle at one point $\textbf{x} = (x, t)$ in space-time to
the conjugate field operator at another point $\textbf{x}'= (x',
t')$:
\begin{eqnarray}
G_{- +}(\textbf{x}, \textbf{x}') & = & ~~i \bra
\Psi^{\dag}(\textbf{x}')\Psi( \textbf{x})\ket ,~~~~ t \in C_- ,~~
t' \in C_+ \nonumber \\ G_{+ -}(\textbf{x}, \textbf{x}') & = & -i
\bra \Psi(\textbf{x})\Psi^{\dag}(\textbf{x}')\ket,~~~~t \in C_+,~
t' \in C_- \nonumber
\\ G_{- -}(\textbf{x}, \textbf{x}') & = & -i \bra T \Psi(\textbf{x})\Psi^{\dag}( \textbf{x}')\ket,
~~ t, t'~~\in C_- \nonumber \\ G_{+ +}(\textbf{x}, \textbf{x}') &
= & -i \bra \widetilde{T}\Psi(\textbf{x})\Psi^{\dag}(
\textbf{x}')\ket, ~~ t, t'~~\in C_+
\end{eqnarray} where $T$ and $\widetilde{T}$ denote the time and anti-time
ordering operations respectively, and $C_+$ and $C_-$ are the time
integration paths mentioned above, i.e. they are the two branches
of the loop C. The subscripts for the Green's functions denote the
branches of the contour in the complex time plane on which the
appropriate time coordinates reside.

After this sketch of the CTP method, our first task is to show how
this powerful formulation can be incorporated in the path-integral
approach to bosonization based on decoupling changes in the
functional integration measure \cite{Nosotros}. To this end we
consider a non covariant version of the Thirring model which has
been previously used to describe the low-energy physics of a
system of non relativistic 1D electrons. In order to illustrate
the way in which a non equilibrium situation can be managed in our
approach, we add an interaction that depends explicitly on time.
This problem gives us the opportunity to go beyond a purely
academic computation by choosing a type of perturbation that
involves interesting physical effects. This is the case of dynamic
barriers that can be also viewed as time-dependent spatially
localized impurities \cite{Komnik} \cite{Gefen}.

We start from the Euclidean action given by
\begin{equation}
S = S_{0} + S_{int} + S_{I},\label{a}
\end{equation}
where $S_{0}$ is the unperturbed action (in the condensed matter
context it is thought of as a linearized free dispersion relation
valid to explore the low energy-long distance physics):
\begin{equation}
 S_{0} = \intx\bP i\slp\Psi ,
\end{equation}
$S_{int}$ describes the forward scattering of spinless fermions
(electrons):
\begin{equation}
 S_{int} = -\frac{1}{2} \intxy ( \bP \gamma_{\mu} \Psi )
(\textbf{x})~ V_{(\mu)}(\textbf{x},\textbf{y}) ~( \bP \gamma_{\mu}
\Psi ) (\textbf{y})
\end{equation}
in which the fermionic currents $\bP \gamma_{\mu} \Psi$  are
coupled through bilocal distance-dependent potentials,
$V_{(\mu)}(\textbf{x},\textbf{y})$. In terms of these potentials
one can make direct contact with the forward-scattering sector of
the "g-ology" model currently used to describe different
scattering processes characterized by coupling functions $g_1$,
$g_2$, $g_3$ and $g_4$ \cite{g-ology}. Neglecting processes
associated to large momentum transfers, only the forward
scattering couplings $g_2$ and $g_4$ play a role. The relation
between these strengths and our potentials are given (in Fourier
space) by
\begin{eqnarray}
g_{2}(\textbf{p}) & = \frac{1}{2}(V_{(1)}(\textbf{p}) +
V_{(0)}(\textbf{p})) \nonumber
\\ g_{4}(\textbf{p}) & = \frac{1}{2}(V_{(0)}(\textbf{p}) - V_{(1)}(\textbf{p})).
\end{eqnarray}

Finally, $S_{I}$ describes the interaction between the electrons and
a localized time-dependent perturbation:
\begin{equation}
S_{I} = \intx\bP \,\gamma_{0} \,\Delta (x, t )\,\Psi ,
\end{equation}
where $\gamma_{0}$ is a Dirac matrix and $\Delta (x, t )$ contains
the details of the perturbation, i.e. the way in which the
interaction is switched on in time and the form in which it is
localized in space. We will make a particular choice for this
function later. In all cases, $\int_C$ indicates that the
integration is defined along the time contour that we briefly
described above.

Let us now briefly show that all the steps followed in
\cite{Nosotros}, in order to express the functional integral
associated to (\ref{a}) as a fermionic determinant, are also valid
when the time-dependent perturbation is taken into account. We
consider the vacuum to vacuum functional:
\begin{equation}
{\Z} = \int_C {\D}\bP {\D} \Psi e^{-S}
\end{equation}
and recall that the electron-electron interaction piece of the
action can be written as
\begin{equation}
S_{int} = -\frac{1}{2} \intx  J_{\mu} K_{\mu}
\end{equation}
where $J_{\mu}$ is the usual fermionic current, and $K_{\mu}$ is a
new current defined as
\begin{equation}
K_{\mu} = \inty V_{(\mu)}(\textbf{x},\textbf{y})\,
J_{\mu}(\textbf{y})
\end{equation}

Using a functional delta and introducing auxiliary vector fields
$\tilde{A}_{\mu}$ and $\tilde{B}_{\mu}$ in the path integral
representation of ${\cal Z}$ (See \cite{Nosotros} for details) one
obtains:
\begin{equation}
{\Z} =  {\N} \int_C  {\D}\bP  {\D}\Psi  {\D} \tilde{A}_{\mu} {\D}
\tilde{B}_{\mu}\exp \left\{ -\intx [ \bP( i\slp +  \gamma_0
\Delta(x,t))\Psi + \tilde{A}_{\mu} \tilde{B}_{\mu} +
\frac{1}{\sqrt{2}}(\tilde{A}_{\mu} J_{\mu} + \tilde{B}_{\mu}
K_{\mu}]\right\}\end{equation}
If we define
\begin{eqnarray}
\bar{B}_{\mu}(\textbf{x}) & = & \inty
V_{\mu}(\textbf{y},\textbf{x}) \tilde{B}_{\mu}(\textbf{x})
\\ \tilde{B}_{\mu}(\textbf{x}) & = & \inty
V^{-1}_{\mu}(\textbf{y},\textbf{x})\bar{B}_{\mu}(\textbf{x})
\end{eqnarray}
with $V^{-1}_{\mu}$ satisfying
\begin{equation}
\inty V^{-1}_{\mu}(\textbf{x},\textbf{y})
V_{\mu}(\textbf{z},\textbf{y}) = \delta(x-z) \delta_{C}(t_x -
t_z),
\end{equation}
where $\delta_{C}$ is the delta function extended to the C contour
as:
\begin{equation}
\delta_{C} (t - t') = \left\{
\begin{array}{ll}\delta(t - t') & ~t , t'~~ \mbox{on}~~ C_+\\
- \delta(t - t') & ~ t , t'~~ \mbox{on}~~ C_-\\ 0 & ~\mbox{otherwise}
\end{array} \right.\label{delta}
\end{equation}
and changing auxiliary variables in the form
\begin{eqnarray}
A_{\mu} & = & \frac{1}{\sqrt{2}} (\tilde{A}_{\mu} + \tilde{B}_{\mu})
\\
B_{\mu} & = & \frac{1}{\sqrt{2}} (\tilde{A}_{\mu} - \tilde{B}_{\mu})
\end{eqnarray}
we get
\begin{equation}
{\Z}=  {\N} \int _C {\D}\bP {\D}\Psi {\D} A_{\mu} {\D} B_{\mu}
\exp^{-S(A, B)} \exp \left\{ -\intx [ \bP( i\slp + \gamma_0
\Delta(x,t) -  \lnA)\Psi] \right\}
\end{equation}
where
\begin{equation}
S(A,B) = \frac{1}{2} \intxy
V^{-1}_{\mu}(\textbf{x},\textbf{y})[A_{\mu}(\textbf{x})A_{\mu}(\textbf{y})
- B_{\mu}(\textbf{x})B_{\mu}(\textbf{y})]
\end{equation}

The jacobian associated with the change $(\tilde{A},\tilde{B})
\rightarrow (A,B)$ is field-independent and can then be absorbed
in the renormalization constant ${\N}$. Note also that the field
${B}_{\mu}$ is completely decoupled from both the field
${A}_{\mu}$ and the fermion field and then its contribution (which
describes a negative metric state) can be factorized and absorbed
in ${\N}$. We thus find
\begin{equation}
{\Z} = {\N} \int {\D}A_{\mu} \exp\left(-S[A_{\mu}]\right)\,\det( i
\slp - \lnA + \gamma_0 \Delta)
\end{equation}
where
\begin{equation}
S[A_{\mu}] = \frac{1}{2} \intxy A_{\mu}(\textbf{x})
V_{(\mu)}^{-1}(\textbf{x},\textbf{y}) A_{\mu}(\textbf{y})
\label{SA}
\end{equation}

Having expressed ${\Z}$ in terms of a fermionic determinant we are
ready to apply the machinery of the decoupling approach to
functional bosonization which is based on the following
transformation
\begin{eqnarray}
\Psi(\textbf{x}) & = & e^{ ( \gamma_5 \Phi(\textbf{x})  -i \eta(\textbf{x}) )}\,\chi(\textbf{x}) \\
\bP(\textbf{x}) & = & \bar{\chi}(\textbf{x}) \, e^{(\gamma_5
\Phi(\textbf{x}) + i \eta(\textbf{x}) )}
\\ {\D}\bP {\D}\Psi & = & J(\Phi, \eta) {\D} \bar{\chi} {\D} \chi
\end{eqnarray}
where $\Phi$ and $\eta$ are scalar fields and $J(\Phi, \eta)$ is
the Fujikawa Jacobian \cite{Fujikawa}. As it is well known, the
above transformation permits to decouple the field $A_{\mu}$ from
the fermionic fields if one writes
\begin{equation}
A_{\mu}(\textbf{x}) = \epsilon_{\mu \nu} \partial_{\nu}
\Phi(\textbf{x}) +
\partial_{\mu}\eta(\textbf{x})\label{Amu}
\end{equation}
which can also be considered as a bosonic change of variables with
trivial (field independent) Jacobian. Then, we find
\begin{equation}
\det( i\slp +  \lnA +\gamma_0 \Delta) = J(\Phi,\eta) \det(i\slp +
\gamma_0 \Delta)
\end{equation}

After a suitable regularization the fermionic Jacobian in Fourier
space ($\textbf{x}\rightarrow \textbf{p}=(p_0,p_1)$) reads
\cite{Chen} \cite{Iucci}
\begin{equation}
J(\Phi, \eta) = \exp \left\{ \frac{1}{\pi}\intp \left[ - p_1^2
\Phi(\textbf{p}) \Phi(-\textbf{p}) + p_1^2 \eta(\textbf{p})
\eta(-\textbf{p}) - 2 p_0 p_1 \Phi(\textbf{p})
\eta(-\textbf{p})\right] \right\}.
\end{equation}

The vacuum to vacuum functional is then expressed as
\begin{equation}
{\Z} = {\N} \int {\D}\Phi {\D}\eta \,\exp^{-S[\Phi, \eta]}
\,J(\Phi, \eta) \det( i\slp + \gamma_0 \Delta)\label{PA}
\end{equation}
where $S[\phi, \eta]$ is obtained when one inserts (\ref{Amu}) in
$S[A_{\mu}]$ (see eq. (\ref{SA})):
\begin{equation}
S[\Phi, \eta] = \intp \left[ \Phi(\textbf{p}) \Phi(-\textbf{p})
{\A}(\textbf{p}) + \eta(\textbf{p}) \eta(-\textbf{p})
{\B}(\textbf{p}) + \Phi(\textbf{p}) \eta(-\textbf{p})
{\C}(\textbf{p})\right]
\end{equation}
with
\begin{eqnarray}
{\A}(\textbf{p}) & = & \frac{1}{2}\left[p_0 ^2 {V}_{(1)}^{-1}
(\textbf{p}) + p_1 ^2
({V}_{(0)}^{-1} (\textbf{p}) + \frac{1}{\pi})\right]\nonumber \\
{\B}(\textbf{p})& = & \frac{1}{2}\left[p_0 ^2 {V}_{(0)}^{-1}
(\textbf{p}) + p_1 ^2
({V}_{(1)}^{-1} (\textbf{p}) - \frac{1}{\pi})\right] \nonumber \\
{\C}(\textbf{p})& = & p_0p_1\left({V}_{(0)}^{-1} (\textbf{p})-
{V}_{(1)}^{-1} (\textbf{p}) + \frac{1}{\pi}\right)\label{ABC}.
\end{eqnarray}
We have then obtained a completely bosonized action for the
non-equilibrium system which can now be used as the starting point
to explore specific time-dependent interactions.

\section{Green's functions}
\setcounter{equation}{0}

We are now ready to compute the Green's functions from eq.
(\ref{PA}), by taking functional derivatives with respect to
suitable sources. It is worthwhile noting that these Green's
functions will be time ordered along the contour C, in the complex
time plane. From now on we will restrict our study to contact
electron-electron interactions in which the couplings $V_0$ and
$V_1$ are constants in momentum space.

Let us now focus our attention on the one-particle fermionic
propagator:
\begin{eqnarray}
G_C(\textbf{x},\textbf{x}') = \bra \Psi(\textbf{x})
\bP(\textbf{x}') \ket = \left(\begin{array}{cl} 0 &
G^{R}_C(\textbf{x},\textbf{x}') \\
\\ G^{L}_C(\textbf{x},\textbf{x}') & 0
\end{array} \right)
\end{eqnarray}
where
\begin{eqnarray}\label{Greens}
G^{R,L}_C =  \left(\begin{array}{cl} G^{R,L}_{++} & G^{R,L}_{+-}
\\ \\ G^{R,L}_{-+} & G^{R,L}_{--} \end{array} \right)
\end{eqnarray}
here the subscripts for the Green functions denotes the branches
of the contour in the complex time plane on which the time
coordinates reside. The subscripts $+$ and $-$ refer to fields
defined in the upper and lower branches, respectively,
corresponding to forward $(+)$ and backward $(-)$ time evolution.

Using the decoupling technique described in the previous section
we can factorize the Green's function as a product of a purely
fermionic Green's function and a bosonic part $B_{C}$, exactly as
it happens in the equilibrium case \cite{Nosotros}. To be specific
we shall consider the propagator corresponding to right-moving
fermions (the left-moving component can be obtained in a
completely similar way):
\begin{equation}
G^{R}_{C}(x, t; x^{'}, t^{'}) = G^R_{I,C}(x, t; x^{'},
t^{'})\,B^R_{C}(x - x^{'}, t - t^{'}),\label{A}
\end{equation}
with
\begin{equation}
B^R_{C}(x - x^{'},t - t^{'}) = exp\left\{-\frac{1}{\pi} \int d^2p~
\sin^2 \left[\frac{p_{0}(t - t^{'}) + p_{1}(x - x^{'})}{2}\right]
\frac{{\B}(\textbf{p}) - {\A}(\textbf{p}) + i
{\C}(\textbf{p})}{{\E}(\textbf{p})}\right\},
\end{equation}
where ${\A}$, ${\B}$ and ${\C}$ is given by eq.(\ref{ABC}) and
${\E} = {\C}^2 - 4 {\A} {\B}$. Please note that we are now using a
more explicit notation in the arguments of functions
($\textbf{x}=(x,t)$). Since we are dealing with perturbations that
depend explicitly on space and time, up to now in an arbitrary
way, the Green's functions are not expected to be translationally
invariant neither in time nor in space.

It is convenient to express $B^R_C(x,t)$ in the form:
\begin{equation}
B^R_C(\textbf{x}) = \exp \left(I_1(\textbf{x}) +
I_2(\textbf{x})\right) \label{I}
\end{equation}
where $I_{1,2}$ are given by
\begin{equation}
I_1(\textbf{x}) = \frac{1}{\pi^2} \frac{b-a}{2ab} \int d^2p
\sin^2[\frac{1}{2}\textbf{p}.\textbf{x}]\exp^{-\Lambda
|p_1|}\frac{(p^2_0 +\frac{(a-b-\frac{2}{\pi})}{b-a}p^2_1)}{(p^2_0
+ p^2_1)(p^2_0 + v^2 p^2_1)}
\end{equation}
\begin{equation}
I_2(\textbf{x}) = \frac{1}{\pi^2} \frac{b-a + \frac{1}{\pi}}{a b}
\int d^2p \sin^2[\frac{1}{2}\textbf{p}.\textbf{x}]e^{-\Lambda
|p_1|}\frac{i p_0 p_1}{(p^2_0 + p^2_1)(p^2_0 + v^2 p^2_1)}
\end{equation}
where we have introduce an ultraviolet cutoff $\Lambda$ and two
parameters a and b defined by:
\begin{eqnarray}
a^{-1} & = & V_{(1)} = (g_2 - g_4) \nonumber \\ b^{-1} & = &
V_{(0)} = (g_2 + g_4)\nonumber.
\end{eqnarray}

The integrals $I_{1,2}$ can be explicitly evaluated and one
obtains
\begin{eqnarray}
I_1(\textbf{x})  & = & \frac{1}{2} \ln \left[ \frac{x^2 + (|t| +
\Lambda)^2}{\Lambda^2}\right] - \frac{1}{4} (K + K^{-1})\ln
\left[\frac{x^2 +( v  |t| + \Lambda)^2}{\Lambda^2}\right]\nonumber\\
 I_2(\textbf{x})  & = &  \frac{1}{2} \mbox{sgn} (t) \left(\ln \left[ \frac{\Lambda
+ \mid t\mid -i x}{\Lambda + |t| +i x}\right] - \ln \left[
\frac{\Lambda + v |t| -i x}{\Lambda + v |t| +i x}\right]\right)
\label{12}
\end{eqnarray}
where
\begin{eqnarray} K & = & \sqrt{(1-\frac{1}{\pi} \hat{V}_{(1)})(1+\frac{1}{\pi} \hat{V}_{(0)})}=
\sqrt{\frac{1+ g_{4}/ \pi -g_{2}/ \pi}{1+ g_{4}/ \pi + g_{2}/
\pi}}\nonumber \\ v & = &
\sqrt{(1+\frac{\hat{V}_{(0)}}{\pi})(1-\frac{\hat{V}_{(1)}}{\pi})}
=\sqrt{(1+ \frac{g_2}{\pi} +\frac{g_4}{\pi})(1- \frac{g_2}{\pi}
+\frac{g_4}{\pi})} \nonumber \\
\end{eqnarray}
are the parameters usually defined in the field theory description of
Luttinger liquids. In particular $v$ is the velocity of the
charge-density modes that dominate the low energy physics of these
systems.

Inserting eq.(\ref{12}) in eq.(\ref{I}) we finally obtain
\begin{equation}
B^R_{C}(x, t) = \Theta(t)\frac{(\Lambda + t - i x)}{(\Lambda + v
t - i x)}\frac{\Lambda^{2 \gamma}}{(x^{2} + (v t + \Lambda
)^{2})^{\gamma}} +  \Theta(-t)\frac{(\Lambda - t + i x)}{(\Lambda - v
t + i x)}\frac{\Lambda^{2 \gamma}}{(x^{2} + (-v t + \Lambda
)^{2})^{\gamma}} \label{B}
\end{equation}
where
\begin{equation}
\gamma  =  \frac{1}{4} ( K + K^{-1} - 2)
\end{equation}

Now we consider the fermionic piece $G_{I, C}(x,t; x', t')$ that
satisfies the Euclidean equation:
\begin{equation} \{i \partial_{t} + \partial_{x} + \Delta(x, t ) \} G_{I, C}(x,t; x', t') = -\delta(x - x') \delta_{C}( t - t').
\end{equation}
with $\delta_{C}$ given by (\ref{delta}).

For later convenience, from now on we shall work in Minkowski space
($t \rightarrow it$ and $G \rightarrow -G$). At this point we also
choose a particular form for the perturbation:
\begin{equation}\label{perturbation}
\Delta(x, t ) = \lambda \, \sin(\Omega t )\, \Theta( t )[ \Theta(
x + a/2 ) - \Theta( x - a/2 )]
\end{equation}
where $\Theta(t)$ is the Heaviside function. We then have a separable
harmonic perturbation adiabatically switched on, consisting of a
square barrier of width $a$ which height $\lambda$ oscillates in time
with frequency $\Omega$. A similar situation was recently considered
by Komnik and Gogolin in their study of transport through a tunneling
junction between a metallic lead and a Luttinger liquid
\cite{Komnik}. However two important differences should be stressed.
First of all they did not include the effect of the adiabatic
switching. Secondly, they did not specify a particular form for the
spatial piece of the interaction, but they were able to expressed
time-averaged quantities in terms of a parameter related to the
strength of the interaction. As we shall see, by specifying the
spatial dependence of the barrier we will be able to give explicit
results for the TED, before a temporal mean value is considered.

The Green's function $G_{I, C}(x,t; x', t')$ satisfies:
\begin{equation}
\{ \partial_{t} + \partial_{x} + \lambda \sin(\Omega t ) \Theta( t
)[ \Theta( x + a/2 ) - \Theta( x - a/2 )] \} G_{I, C}(x,t; x', t')
= -\delta ( x - x' ) \delta_{C} ( t - t').
\end{equation}

$G_{I, C}(x,t; x', t')$ can be factorized in terms of the free
(translationally invariant) Green's function and a function to be
determined that contains the effect of the broken translational
symmetry:
\begin{equation}
G_{I, C}(x,t; x', t') = G_{0, C}(x - x', t - t') \exp [ \beta (x,
t) - \beta (x', t')],\label{C}
\end{equation}
where $G_{0, C}(x - x', t - t') $ satisfies:
\begin{equation}
\{ \partial_{t} + \partial_{x}\}G_{0, C}(x - y, t - t^{'}) =
-\delta ( x - y ) \delta_C ( t - t^{'}).
\end{equation}
The solution reads
\begin{equation}
G_{0, C}(x - y, t - t^{'}) = \frac{ \Theta_C( t - t^{'})}{2 \pi
}\frac{1}{[ ( x - y) - ( t - t^{'}) + i\alpha_{0} ]} +
\frac{\Theta_C( t^{'} - t)}{2 \pi}\frac{1}{[ ( x - y) - ( t -
t^{'}) - i\alpha_{0} ]},\label{D}
\end{equation}
where $\alpha_{0}$ is an infrared cut-off, and we have defined the
Heaviside function on the contour in the usual way:
\begin{equation}
\Theta_c (t - t') = \left\{
\begin{array}{ll}\Theta(t - t') & ~t , t'~~ \mbox{on}~~ C_+\\
\Theta(t' - t) & ~ t , t'~~ \mbox{on}~~ C_-\\ 0 & ~ t ~~\mbox{on}
~~C_+ ,~~ t'~~ \mbox{on}~~ C_-
\\ 1 & ~ t' ~~\mbox{on}~~ C_+, ~~ t ~~ \mbox{on}~~ C_- \end{array} \right.
\end{equation}

The function $\beta (x, t)$ obeys the following equation:
\begin{equation}
\{ \partial_{0} + \partial_{1}\}\beta (x, t) = -
\lambda \,\sin(\Omega t )\, \Theta( t )\,[ \Theta( x + a/2 ) - \Theta( x -
a/2 ) ],
\end{equation}
which solution is given by
\begin{equation}
\beta (x, t) = \frac{ \lambda }{2 \pi }\int^{\infty}_{0} d
t^{'} \sin(\Omega t^{'} )\int^{\infty}_{-\infty} d x^{'} [ \Theta(
x^{'} + a/2 ) - \Theta( x^{'} - a/2 ) ]G_{0}(x - x^{'}, t -
t^{'}),
\end{equation}
where $G_{0}$ is the inverse of the $(\partial_{t} +
\partial_{x})$ operator.

We thus obtain
\begin{multline}
\beta (x, t) = -i \frac{\lambda}{\Omega}\Theta ( t)\{ \Theta( - (
x + a/2 )) \cos(\Omega(t - ( x + a/2))) - \Theta( - ( x - a/2 ))
\cos(\Omega(t - ( x - a/2))) + \\ + \cos(\Omega t)[ \Theta( x +
a/2 ) - \Theta( x - a/2 ) ]\}. \label{E}
\end{multline}

At this point we already have all the ingredients needed to build the
Green's function of equation (\ref{A}). Indeed, using (\ref{D}) and
(\ref{E}) in (\ref{C}) we have the first factor in (\ref{A}), which
together with (\ref{B}) give us the complete Green's function. But
before we write down the final result two important observations are
in order. First we note that both (\ref{B}) and (\ref{D}) are
translationally invariant. The effect of the time dependent
interaction is completely contained in the exponential factors. It is
then natural to reexpress $G_{C}$ in the form
\begin{equation}\label{finalG}
G_{C}(x,t; x',  t') = G_{C}^{\gamma}(x - x', t - t^{'}) \exp [
\beta (x, t) - \beta (x', t')],
\end{equation}
where $G_{C}^{\gamma}(x - x', t - t')$ is the product of (\ref{B})
and (\ref{D}), which is nothing but the equilibrium propagator for
a Luttinger liquid with its characteristic interaction dependent
exponent $\gamma$. The second observation concerns the actual form
of $G_{C}^{\gamma}$, which up to this point is the exact
electronic propagator. As it is well-known, even in the
equilibrium case, computations of physical quantities with this
propagator are a non trivial challenge \cite{Voit}, \cite{Meden}.
Therefore we shall consider the usual long-distance, and long-time
approximation $|x|, |t| \gg \Lambda$ which leads to
\begin{multline}
G_{C}^{\gamma}(x - x', t - t') = \\ = \frac{\Theta_{c}( t - t^{'}
)}{2 \pi}\left \{\frac{1}{ v( t - t^{'})- (x - x') -
i\alpha_{0}}\left[\frac{\Lambda^{2}}{( v( t - t')-(x -x') -
i\Lambda)(v( t - t')+ ( x - x') - i\Lambda)}\right]^{\gamma}\right
\} +
\\ + \frac{\Theta_{c}( t' - t )}{2 \pi}\left\{\frac{1}{v( t -
t') -(x -y) + i\alpha_{0}}\left[\frac{\Lambda^{2}}{(v( t - t')- (
x - x') + i\Lambda)( v( t - t') +( x -x') +
i\Lambda)}\right]^{\gamma}\right\}.\label{Ggama}
\end{multline}
Equation (\ref{finalG}), with the explicit forms for
$\beta(\textbf{x})$ and $G_{C}^{\gamma}$ given in (\ref{E}) and
(\ref{Ggama}), is our first non trivial result. We have obtained
an exact (within the long-distance approximation) analytical
expression for the Green's function of a Luttinger liquid in the
presence of a dynamic barrier. As we shall see in the next
section, this result will enable us to compute the energy
distribution of this system.

\section{Total Energy Distribution Function}
\setcounter{equation}{0}

In this section we use the Wigner representation in which the
distribution function is defined in terms of  $G_{-
+}(\textbf{x},\textbf{x}')$. In this representation it is usually
convenient to define a center of mass coordinate system $(R,T) =
(\textbf{x} + \textbf{x}')/2$, $(r,t) = (\textbf{x} -
\textbf{x}')$. One then Fourier transforms the relative variable,
and in this way, in terms of the distribution function one obtains
macroscopic quantities such as the particle density, particle
current, energy density and energy current. Here we are specially
interested in computing the total energy distribution (TED) of the
electrons in the presence of the time-dependent perturbation
defined in eq.(\ref{perturbation}) . The TED is defined as
\begin{equation}
n(\omega, x, t) = - i \int d \tau \exp ( i \omega \tau )\,G_{-
+}(x, t + \tau/2;x, t - \tau/2),
\end{equation}
where we now call $p_0=\omega$. Of course, we can use the results
obtained in the previous section for the Green's functions, in
order to compute $n(\omega, x, t)$ (see eq. (\ref{Greens})). As
before, we shall consider the contribution to the TED of the
right-moving particles only ($n_R(\omega, x, t)$. When one tries
to obtain an explicit expression for this quantity, it becomes
apparent that the necessary stages of the calculation are greatly
simplified if one considers the large-distance (time)
approximation $|x|, |t| \gg \Lambda$. As a first step, for
illustrative purposes we will evaluate, within this approximation,
the TED to first order in $\lambda/\Omega$ (computation of higher
orders is tedious but straightforward). Thus we write
\begin{equation}
n_R(\omega, x, t) = n_{0}(\omega) - i\frac{\lambda}{\Omega} n_{I}
(\omega, x, t),
\end{equation}
where the first term in the right hand side is the equilibrium TED
given by
\begin{equation}
n_0(\omega) = \frac{\Lambda^{2 \gamma}}{2 \pi} \int d \tau \frac{
\exp (i \omega \tau)}{(\alpha_0 + i v \tau)( \Lambda + i v \tau)^{2
\gamma}} \label{F}
\end{equation}
which, after taking the limit $\alpha_0\rightarrow0$, yields
\begin{equation}
n_{0}(\omega) = \frac{\exp\left(- \Lambda \omega/v\right)}{v
\Gamma ( 1 + 2\gamma )}\left( \frac{ \Lambda \omega}{v}\right)^{2
\gamma}\Phi \left( 1 , 1 + 2 \gamma ; \frac{ \Lambda \omega}{v}
\right)\Theta (\omega),
\end{equation}
where $\Phi ( a , b; z)$ is Kummer's confluent hypergeometric
function. This result agrees with previous calculations
\cite{Voit},\cite{Meden}. In particular, in the low-frequency limit
we reobtain the expected behaviour $n_{0}(\omega)\sim
\omega^{2\gamma}$.

The first non trivial correction is
\begin{eqnarray}
n_I( \omega, x, t) & = & \frac{\Lambda^{2 \gamma}}{4 \pi}  \{ \int_{-2t}^{\infty}\frac{ d
\tau}{ (\alpha_0 + i v \tau)(\Lambda + i v \tau)^{2 \gamma}} [ \exp
(i (\omega + \Omega/2)\tau)\exp(i \Omega t) F(x,a,\Omega) \nonumber
\\ & + &  \exp
(i (\omega - \Omega/2)\tau)\exp( - i \Omega t) F(x,a,- \Omega)]
\nonumber \\ & - & \int_{-\infty}^{2 t}\frac{ d \tau}{ (\alpha_0 +
i v \tau)(\Lambda + i v \tau)^{2 \gamma}} [ \exp (i (\omega -
\Omega/2)\tau)\exp(i \Omega t) F(x,a,\Omega) \nonumber
\\ & + &  \exp
(i (\omega + \Omega/2)\tau)\exp( - i \Omega t) F(x,a,- \Omega)]\}
\end{eqnarray}
which can be cast in the form
\begin{eqnarray}
n_I( \omega, x, t) & = & \frac{\Lambda^{2 \gamma}}{4 \pi}
 [ \exp( i\Omega t) F(x, a, \Omega)
 \int_{-2t}^{\infty}\frac{ d
\tau  \exp (i (\omega + \Omega/2)\tau)}{ (\alpha_0 + i v
\tau)(\Lambda + i v \tau)^{2 \gamma}} + \nonumber \\ & + & \exp( -
i\Omega t) F(x, a, - \Omega)
 \int_{-2t}^{\infty}\frac{ d
\tau  \exp (i (\omega - \Omega/2)\tau)}{ (\alpha_0 + i v
\tau)(\Lambda + i v \tau)^{2 \gamma}} + \nonumber \\ & - & \exp(
i\Omega t) F(x, a, \Omega)
 \int_{-\infty}^{2t}\frac{ d
\tau  \exp (i (\omega - \Omega/2)\tau)}{ (\alpha_0 + i v
\tau)(\Lambda + i v \tau)^{2 \gamma}} + \nonumber \\ & - & \exp( -
i\Omega t) F(x, a,-\Omega)
 \int_{-\infty}^{2 t }\frac{ d
\tau  \exp (i (\omega + \Omega/2)\tau)}{ (\alpha_0 + i v
\tau)(\Lambda + i v \tau)^{2 \gamma}} ] ,\label{G}
\end{eqnarray}
where
\begin{eqnarray}\label{functionF}
F( x , a ,\Omega)& = \Theta ( - ( x + a/2) )\exp [ - i \Omega
( x + a/2 ) ] - \Theta ( - ( x - a/2) )\exp [ - i \Omega ( x - a/2 )
] + \nonumber \\& +  \Theta ( x + a/2 ) - \Theta (  x - a/2
).
\end{eqnarray}

At this point we consider two important limits: for very large
times in the past ($t \rightarrow - \infty$) one obtains $ n_I =
0$, i.e. the TED recovers its equilibrium value, as expected. In
the other case, for large values of t in the future, the integrals
in the above expression coincide with (\ref{F}) and one has:
\begin{equation}
n_I(\omega, x, t) = \frac{1}{2} [ \exp( i \Omega t) F(x, a,
\Omega) - \exp(-i \Omega t) F(x,a, -\Omega)][n_0(\omega +\Omega/2)
- n_0(\omega - \Omega/2)].
\end{equation}

Then, up to first order in $\lambda$, for $t \rightarrow -\infty$
\begin{equation}
n_R(\omega, x,t) = n_0(\omega),
\end{equation}
and for very large positive times:
\begin{equation}
n_R(\omega, x, t) = n_0(\omega) - \frac{i \lambda}{2 \Omega} [
\exp(i \Omega t) F(x, a, \Omega) - \exp(-i \Omega t) F(x,a,
-\Omega)][n_0(\omega + \Omega/2) - n_0(\omega - \Omega/2)].
\end{equation}

Thus, in this case the non-equilibrium correction to the TED is a
superposition of unperturbed TED's centered at energies $\pm
\Omega/2$ .

In the large time approximation we were able to extend the above
result to all orders in $\lambda$. Indeed, taking into account
that, in this regime one can write
\begin{eqnarray}
\exp ( \beta(x, t + \tau/2)) - \exp(\beta(x, t - \tau/2)) &
\approx & \sum_{n=0}^{\infty} \left(\frac{-i \lambda}{2
\Omega}\right)^n \frac{1}{n!} \left(\exp(i \Omega \tau/2) - \exp(
-i \Omega \tau/2)\right)^n \nonumber \\ & \times & \left( \exp(i
\Omega t) F(x,a, \Omega) - \exp(-i \Omega t)
F(x,a,-\Omega)\right)^n \nonumber,
\end{eqnarray}
and using the binomial expansion
\begin{equation}
\left(\exp(i \Omega \tau/2) - \exp( -i \Omega \tau/2)\right)^n =
\sum_{k = 0}^{n} \left( \begin{array}{c} n \\ k  \end{array}
\right)(-1)^{n -k} \exp(i \Omega \tau ( n - 2 k)/2),
\end{equation}
one finally gets
\begin{eqnarray}\label{functionn_R}
n_R(\omega, x, t)  =  \sum_{n=0}^{\infty} \sum_{k
=0}^{n}\,\frac{1}{n!}\,\left( \frac{-\lambda i}{2
\Omega}\left(\exp(i \Omega t) F(x,a,\Omega) - \exp( -i \Omega t)
F(x,a - \Omega)\right)\right)^n  \times \nonumber \\  \left(
\begin{array}{c} n
\\ k
\end{array}\right) (-1)^{n -k} n_0(\omega + \Omega(n-2 k)/2),
\end{eqnarray}
which is one of the main results of this work. We derived, in the
large time regime, an analytical expression for the TED at all
orders in $\lambda/\Omega$.

In realistic systems the frequency $\Omega$ is expected to be
quite high so that it is unlikely that the explicit time
resolution of the TED would be experimentally accessible. Then, it
is natural to consider the time average, over the period of the
perturbation, for our TED: $<n_R(\omega, x)>\equiv N(\omega, x)$.
In so doing one readily discovers that only even powers of
$\lambda/\Omega$ survive the mean value:
\begin{equation}\label{functionN}
N(\omega,x) = \sum_{n=0}^{\infty} \sum_{k =0}^{2 n}\left(
\begin{array}{c} 2 n
\\ n
\end{array}\right) \frac{(-1)^{2 n -k}}{(2 n - k)!k!}\left( \left(
\frac{\lambda}{2 \Omega}\right)^{2} F(x,a, \Omega) F(x,a,
-\Omega)\right)^{n} n_0(\omega + \Omega (n - k))
\end{equation}

Then, the averaged TED is a superposition of unperturbed TED's
centered at energies $n\,\Omega$, with n integer. The terms in
this linear combination are weighted by powers of
$z=(\frac{\lambda}{2 \Omega})^{2} F(x,a, \Omega) F(x,a, -\Omega)$,
i.e. factors associated to the strength of the interaction and its
geometry. The interpretation of the physical consequences of the
combined effect of the electron-electron interaction and the time
dependent impurity is facilitated by evaluating $N(\omega,x)$ in a
particular spatial point and considering the function
$N(\omega,z)$, as was done in \cite{Komnik}. It should be
stressed, however, that our $z$-parameter differs from the one
defined in that work, in which the absence of adiabatic switching
allowed the authors to compute $N(\omega,z)$ without specifying a
particular spatial dependence for the scatterer.

\begin{figure}
\begin{center}
\includegraphics{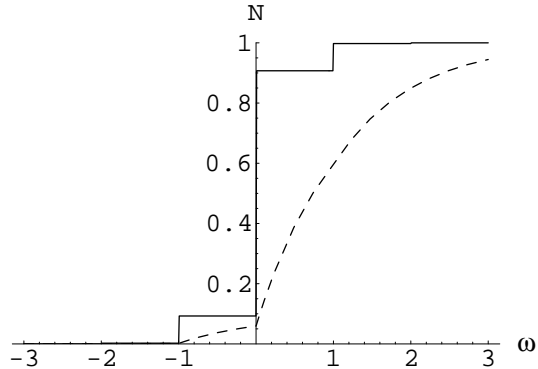}
\caption{\label{fig1ne}: $N(\omega,z=0.1)$ for $\gamma=0$ (filled
line) and $\gamma=1/2$ (dashed line).}
\end{center}
\end{figure}

\begin{figure}
\begin{center}
\includegraphics{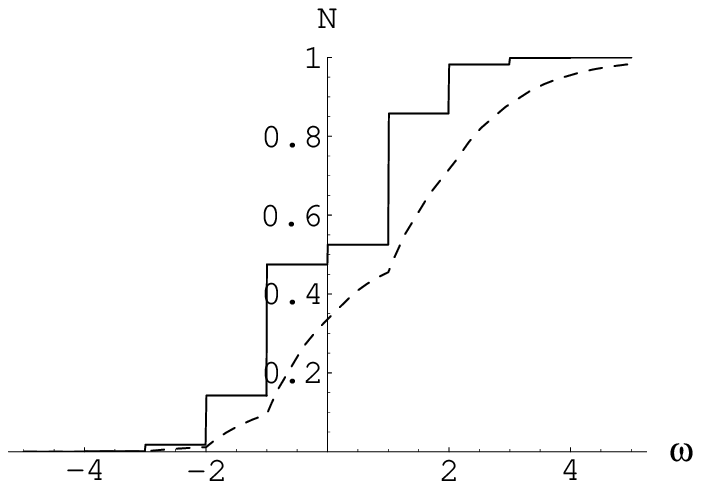}
\caption{\label{fig2ne}: $N(\omega,z=1)$ for $\gamma=0$ (filled
line) and $\gamma=1/2$ (dashed line).}
\end{center}
\end{figure}

\begin{figure}
\begin{center}
\includegraphics{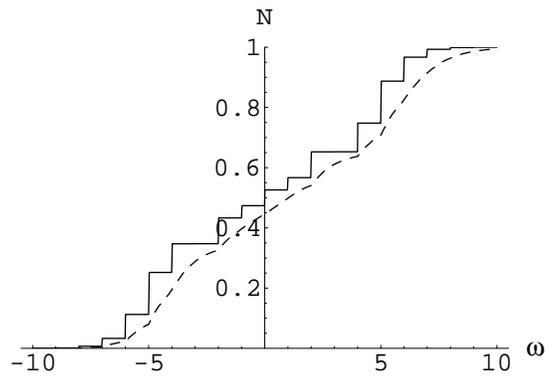}
\caption{\label{fig3ne}: $N(\omega,z=10)$ for $\gamma=0$ (filled
line) and $\gamma=1/2$ (dashed line).}
\end{center}
\end{figure}

In figures 1, 2 and 3, we show the behaviour of the function
$N(\omega,z)$ as function of $\omega$, for $z=0.1, 1$ and $10$,
respectively. We have set $\Omega=v=\Lambda=1$, and recall that
the Fermi energy has been put equal to zero. In each figure we
have considered the effect of the dynamic barrier for both the
free ($\gamma=0$, filled line) and interacting ($\gamma=1/2$,
dashed line) cases. The results are in qualitative agreement with
the ones obtained in \cite{Komnik} for a generic time-dependent
impurity without adiabatic switching. First of all one sees the
appearance of equidistant sidebands in the TED corresponding to
the noninteracting system \cite{sidebands}. As it is known, this
is a manifestation of the fact that, in the presence of a time
dependent perturbation, scattering ceases to be elastic and gain
or loss of energy can take place. When forward-scattering is taken
into account the step functions of the free case become smoothed,
and the values of $N(\omega,z)$ are always below the ones
corresponding to the noninteracting case, as expected. For
increasing $z$, the number of sidebands also increases and the
free and interacting values of $N(\omega,z)$ become closer from
each other.

\begin{figure}
\begin{center}
\includegraphics{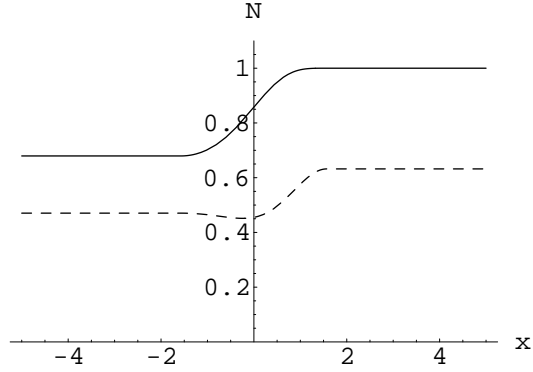}
\caption{\label{fig4}: $N(x)$ for $a=\pi$, $\gamma=0$ (filled
line) and $\gamma=1/2$ (dashed line).}
\end{center}
\end{figure}

\begin{figure}
\begin{center}
\includegraphics{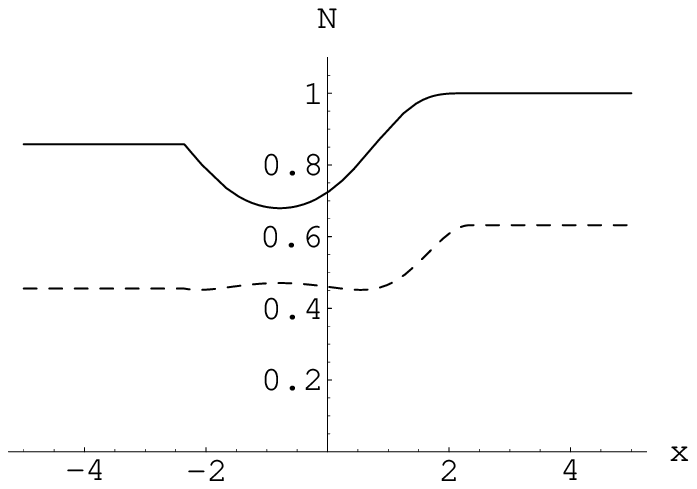}
\caption{\label{fig5}: $N(x)$ for $a=3\pi/2$, $\gamma=0$ (filled
line) and $\gamma=1/2$ (dashed line).}
\end{center}
\end{figure}

\begin{figure}
\begin{center}
\includegraphics{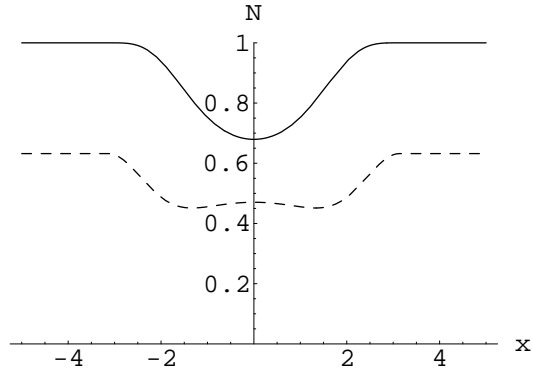}
\caption{\label{fig6}: $N(x)$ for $a=2\pi$, $\gamma=0$ (filled
line) and $\gamma=1/2$ (dashed line).}
\end{center}
\end{figure}

Since we have obtained not only the TED as function of $\omega$,
but also its spatial dependence, we can also analyze the spatial
distribution of energy for given energies. In figures 4, 5 and 6
we plotted $N(x)$ for $\lambda=\omega=1$ and $a=\pi, 3\pi/2$ and
$2\pi$ respectively. As before, the filled line corresponds to the
free case and the dashed line represents the behavior for
$\gamma=1/2$. Due to the specific form of the impurity, the
modulating factors in (\ref{functionN}), which are determined by
the function $F(x,a,\Omega)$ (see equation (\ref{functionF})),
vanish for $x>a/2$ (please remember that we are considering the
time averaged TED for the right movers only). For this reason
$N(x)$ acquires the unperturbed (equilibrium) value to the right
of the barrier. On the other hand, the behavior of $N(x)$ to the
left of the scatterer is governed by the value of
$\sin^2(\Omega\,a/2)$. This is why when we vary $a$ from
$\pi/\Omega$ to $2\pi/\Omega$ the modulation decreases until it
reaches the zero value, for $a=2\pi/\Omega$. This last case
corresponds to the symmetric distribution represented in figure 6,
which is in fact the case of interest in realistic systems.
Indeed, in carbon nanotubes \cite{nano} and quantum wires
\cite{wires}, the width of the impurities are of the order of
nm´s. Then, recovering the dependence on the Fermi velocity (which
we had set equal to 1, but is of order $8\times10^5$ m/s) and
considering external frequencies in the range of the far
megahertzs, one concludes that $\Omega\,a/(2\,v_F)\approx
10^{-8}$.

Finally, in figures 7 and 8, we present three-dimensional plots of
$N$ as function of both $\omega$ and x.

\begin{figure}
\begin{center}
\includegraphics{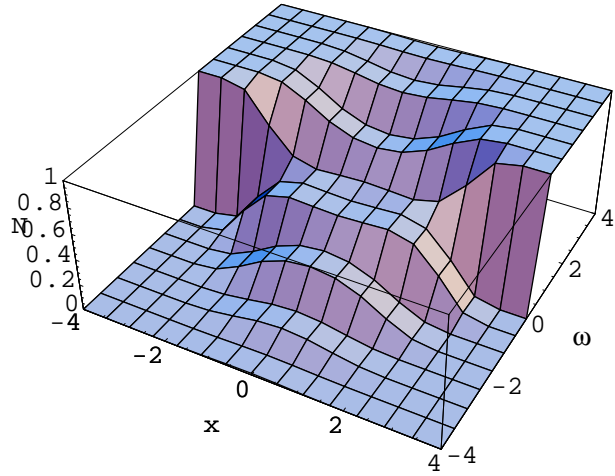}
\caption{\label{fig7}: $N(\omega,x)$ for $a=2\pi$ and $\gamma=0$.}
\end{center}
\end{figure}

\begin{figure}
\begin{center}
\includegraphics{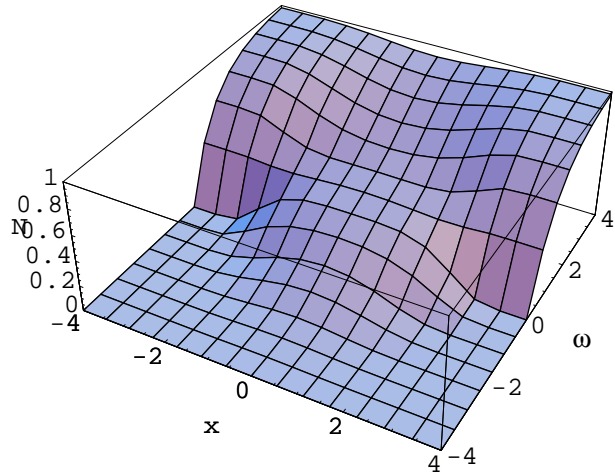}
\caption{\label{fig8}: $N(\omega,x)$ for $a=2\pi$ and
$\gamma=1/2$.}
\end{center}
\end{figure}

\newpage
\section{Conclusions}
We have undertaken the extension of a functional bosonization
method, previously developed in the context of equilibrium field
theory, to the case in which time-dependent perturbations are
included. In particular, we have considered a model of fermions
interacting through current-current interactions. In order to make
contact with the description of 1D condensed matter systems those
interactions are incorporated in a non-covariant way. In addition
to this, by now standard (equilibrium) setting, we also included a
term in the action -bilinear in fermion fields- which describes
scattering through a time-dependent barrier. By using the CTP
(Schwinger-Keldysh) approach to manage the time evolution of a
field theory, we obtained the exact fermionic Green's function for
our out of equilibrium Thirring model (equation (\ref{finalG})).
In the long-distance and long-time regime (the usual situation
considered in the condensed matter context) we obtained a closed
expression for the TED (equation (\ref{functionn_R})), which is a
superposition of equilibrium TED's centered at energies equal to
integer multiples of the external half-frequency $\Omega/2$ . From
this expression we readily obtained the time averaged TED
$N(\omega,x)$ (equation (\ref{functionN})), in which only
contributions corresponding to integer multiples of $\Omega$
survive, modulated by factors related to the spatial distribution
of the impurity. Our results are consistent with the findings of
the authors of ref. \cite{Komnik}, who computed the temporal mean
value of the TED for an arbitrary impurity geometry in a specific
spatial point. In contrast to these authors, we included a
temporal Heaviside function in the time dependent interaction.
Although for this case we could not compute the TED for arbitrary
scatterer geometries, by considering a rectangular barrier we were
able to obtain explicit analytical results for the time-averaged
TED, not only as function of energy but also in terms of the
spatial coordinate.

This work could be continued in several directions. Firstly, it
would be desirable to include the effect of finite temperature.
Secondly, one could try to explore the short-time regime by
building the TED using directly equation (\ref{finalG}) for the
Green's function instead of the long-time approximation resulting
through the simplified expression for $G_{C}^{\gamma}$ (equation
(\ref{Ggama})). This does not seem to be an easy task, but
nevertheless we hope to follow this route in the close future.

\section*{Acknowledgements}
We thank Andrei Komnik for valuable e-mail exchanges. We are grateful to Silvana Stewart for helpful comments.\\
This work was partially supported by Universidad Nacional de La
Plata  and Consejo Nacional de Investigaciones Cient\'{\i}ficas y
T\'ecnicas, CONICET (Argentina).

\end{document}